\documentclass[12pt,preprint]{aastex}
\usepackage{psfig}

\newcommand\mzon   {M$_{\odot}$}
\newcommand\pp     {$\pm$}

\def\degr{\hbox{$^\circ$}}
\newcommand\Lunit   {erg s$^{-1}$}
\newcommand\funit   {erg cm$^{-2}$ s$^{-1}$}

\righthead{X 1732--304 in quiescence}
\slugcomment{Accepted for publication in ApJ Main Journal, 20 February
2002}

\begin{document}

\title{A {\itshape Chandra} observation of the globular cluster Terzan
1: the neutron-star X-ray transient X 1732--304 in quiescence}

\author{Rudy Wijnands\altaffilmark{1,3}, Craig
O. Heinke\altaffilmark{2}, Jonathan E. Grindlay\altaffilmark{2}}

\altaffiltext{1}{Center for Space Research, Massachusetts Institute of
Technology, 77 Massachusetts Avenue, Cambridge, MA 02139-4307, USA;
rudy@space.mit.edu}

\altaffiltext{2}{Harvard College Observatory, 60 Garden St.,
Cambridge, MA 02138}

\altaffiltext{3}{Chandra Fellow}

\begin{abstract}
We present a short ($\sim$3.6 ks) {\it Chandra}/HRC-I observation of
the globular cluster Terzan 1. This cluster is known to contain the
bright neutron star low-mass X-ray binary X 1732--304 which was active
during the 1980's and most of the 1990's. But a {\it BeppoSAX}
observation performed in 1999 only showed a very weak source
indicating that the source had become quiescent.  During our {\it
Chandra} observation, we detect one source with a 0.5--10 keV
luminosity of approximately $1 - 2 \times10^{33}$ \Lunit~(for an
assumed distance of 5.2 kpc). However, its position is not consistent
with that of X 1732--304.  We do not conclusively detect X 1732--304
with a 0.5--10 keV luminosity upper limit of $0.5 - 1\times10^{33}$
\Lunit. This limit is consistent with the luminosities observed for
several neutron-star X-ray transients in our Galaxy when they are
quiescent, strongly suggesting that X 1732--304 was still quiescent
during our {\it Chandra} observation. If the quiescent emission in
neutron star X-ray transients is due to the thermal emission from the
neutron star, then it is expected that the quiescent luminosity
depends on the time-averaged accretion rate of the source. However,
the upper limit on the quiescent luminosity of X 1732--304, combined
with its very long accretion episode prior to the current quiescent
episode, indicates that the quiescent episodes of the source have to
be longer than $\sim$200 years. This would be the second system after
KS 1731--260 for which quiescent episodes longer than several hundreds
of years have been inferred. We discuss this possibility and
alternative quiescent models to explain our results.

\end{abstract}


\section{Introduction \label{section:intro}}

In 1980, {\it Hakucho} detected a bursting X-ray source in the
direction of the globular cluster Terzan 1 (Makishima et al. 1981;
Inoue et al. 1981). Several years later, a steady X-ray source was
detected (X 1732--304) consistent with this globular cluster and it is
most likely the same source as the bursting source (Skinner et
al. 1987; Parmar, Stella, \& Giommi 1989).  Since then, the source has
persistently been detected at 2--10 keV luminosities between a few
times $10^{35}$ erg s$^{-1}$ and $\sim10^{37}$ erg s$^{-1}$ (see
Figure 3 of Guainazzi, Parmar, \& Oosterbroek 1999 and references
therein). The source was detected with the {\it ROSAT} high resolution
imager, which constrained the position of the source to $\sim$5$''$
(Johnston, Verbunt, \& Hasinger 1995). Also, a radio source was
detected with the {\it VLA} in the {\it ROSAT} error circles (Mart\'
\i~ et al. 1998) and it might be the radio counterpart of X
1732--304.

Guainazzi et al. (1999) reported on an anomalous low-state from X
1732--304 during a 1999 {\it BeppoSAX} observation. They could only
detect one dim source with a 2--10 keV luminosity of
$1.9\times10^{33}$ erg s$^{-1}$ (for a distance of 5.2 kpc [Ortolani
et al. 1999]; note that Guainazzi et al. 1999 used 4.5 kpc). This
source luminosity and its X-ray spectrum are both very similar to
those observed for the neutron star transients in the Galaxy when they
are in their quiescent state. These similarities strongly indicate
that X 1732--304 suddenly turned off and became quiescent after having
accreted for more than 12 years. This conclusion also holds when this
{\it BeppoSAX} source is not X 1732--304 but an unrelated source,
likely also part of the globular cluster (Guainazzi et al. 1999). The
long active episode of X 1732--304 makes it very similar to the
neutron star X-ray transient KS 1731--260, which was also active for
more than a decade (see Wijnands et al. 2001, who called such systems
'long-duration' X-ray transients).

Recently, it was realized (Wijnands et al. 2001) that when such
long-duration X-ray transients turn off again, that they could be used
to study the effects of a prolonged period of accretion on the neutron
star core and crust, and on the the quiescent properties of neutron
star X-ray transients.  At the end of 2000 or early 2001, KS 1731--260
suddenly turned off after an accretion episode which lasted for at
least 12 years.  A {\it Chandra} observation on this source performed
in March 2001 (just a few months after this source turned off) showed
that the quiescent luminosity and temperature of this source were very
similar to those of the ordinary transients in quiescence (Wijnands et
al. 2001). The {\it BeppoSAX} observation of KS 1731--260, which was
performed a few weeks before the {\it Chandra} one, showed very
similar source properties (Burderi et al. 2002). If the quiescent
emission in neutron star transients is due to thermal emission from
the neutron star (e.g, van Paradijs et al. 1987), then the exact
quiescent luminosity should depend on the time-averaged accretion rate
of the system (e.g., Campana et al. 1998; Brown, Bildsten, \& Rutledge
1998). If true, then KS 1731--260 has to be in quiescence in between
outbursts for over a thousand years in order for the neutron star to
be as cool as measured (Wijnands et al. 2001; Rutledge et al. 2002).

It is unclear if KS 1731--260 is unique in its behavior or that more
sources behave similarly. Several more long-duration transients are
known and among them, X 1732--304 is one of the best candidates to
study in quiescence and to compare with KS 1731--260, because it is
currently quiescent. Here we report our analysis of a short {\it
Chandra} observation on this source during quiescence.

\section{Observation, analysis, and results}

\begin{figure}
\begin{center}
\begin{tabular}{c}
\psfig{figure=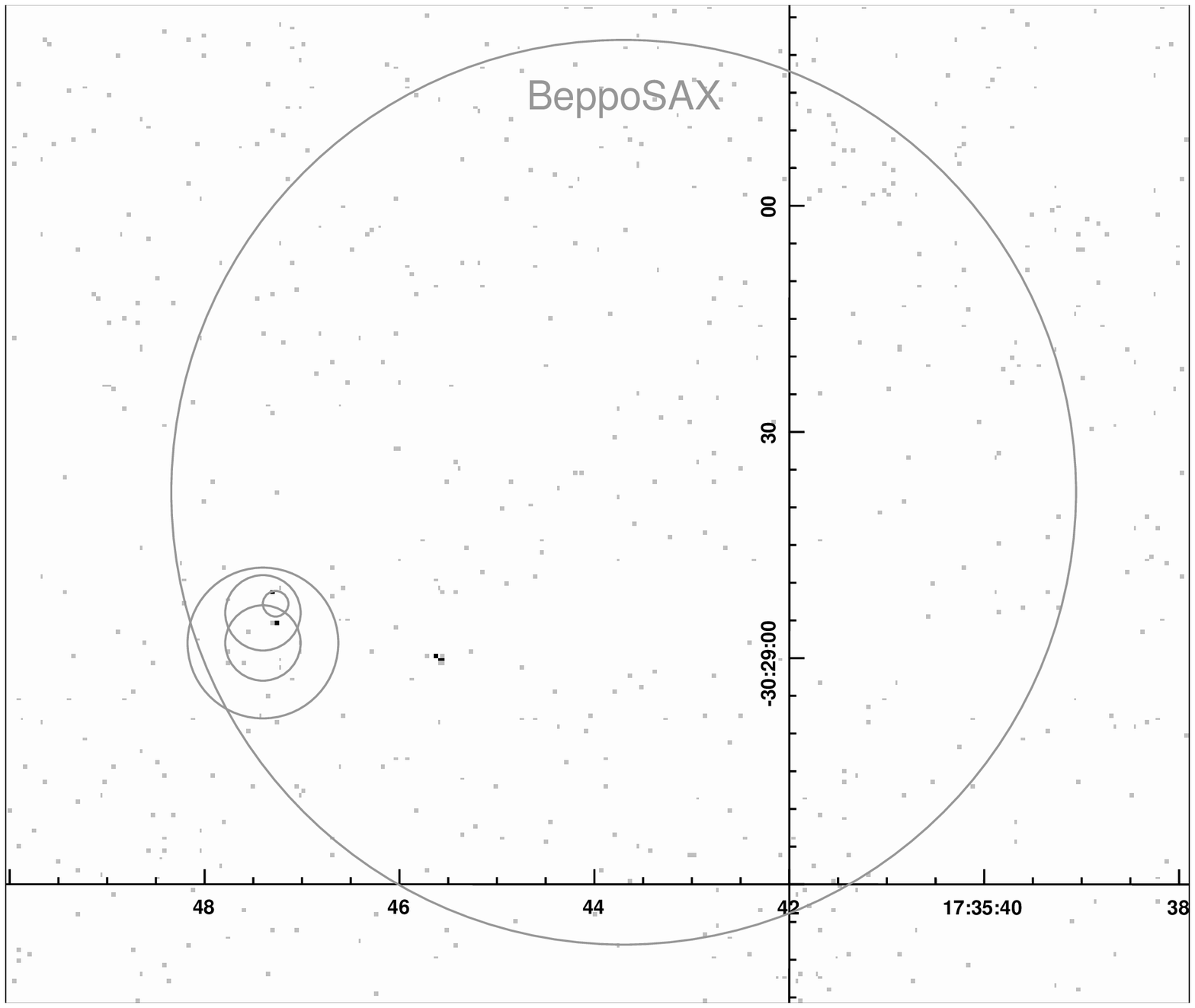,width=8cm}
\psfig{figure=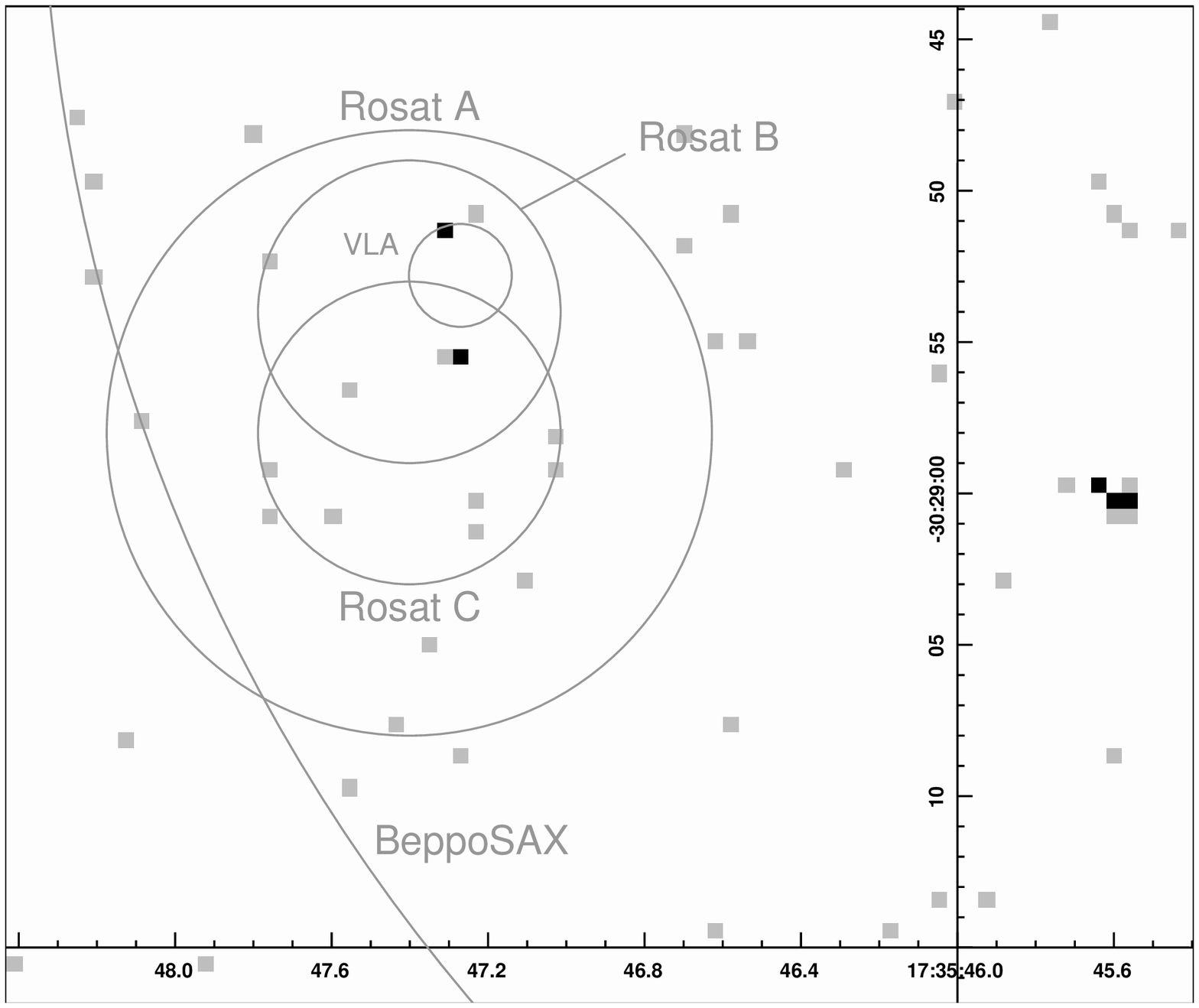,width=8cm}
\end{tabular}
\figcaption{The {\it Chandra}/HRC-I image of the field of Terzan 1,
with a bin size of 0.5$''$. The left panel shows the field covering
the total {\it BeppoSAX} error circle (using a radius of 1$'$;
Guainazzi et al. 1999). The right panel shows a close-up of the field
covering the {\it ROSAT} error circles (see Johnston et al. 1995 for
the three {\it ROSAT} pointings and the lettering of the
observations). Also shown is the error circle of the radio source
detected with the {\it VLA} (Mart\' \i~ et al. 1998).
\label{fig:chandra} }
\end{center}
\end{figure}

The {\it Chandra} observation used in this paper was performed on 9
March 2000 for a total of 3665 seconds of on source time and using the
HRC-I instrument. We used the CIAO tools and the threads listed at
http://asc.harvard.edu to analyze the data.  The central part of the
obtained image is displayed in Figure~\ref{fig:chandra}. In the left
panel, the field of the {\it BeppoSAX} error circle is shown
(Guainazzi et al. 1999) and in the right panel a close up of the {\it
ROSAT} and {\it VLA} error circles (Johnston et al. 1995; Mart\' \i~
et al. 1998). In the HRC-I field containing the {\it BeppoSAX} error
circle only one source was detected using the tool {\it
wavdetect}. The position of this source, as determined with this tool
(R.A = 17$^h$ 35$^m$ 45.603$^s$, Dec. = --30\degr~29$'$ 00.1$''$; all
coordinates in this paper are for J2000.0; the error on the position
is dominated by the pointing accuracy of the satellite and is
typically 0.6$''$; 1$\sigma$, Aldcroft et al. 2000), is inconsistent
with that of the {\it ROSAT} and {\it VLA} positions of X 1732--304
(see Fig.~\ref{fig:chandra}). Therefore, this source cannot be the
quiescent X-ray counterpart of X 1732--304; we designate this source
CXOGLB J173545.6--302900. We used the tool {\it dmextract} to extract
the number of source counts in a 3$''$ circle around the source
position. The number of background counts was estimated by using an
annulus from 3$''$ to 10$''$ around the same position. In total, we
detected only 11 counts from the source position, and according to
{\it dmextract} about 1 count is most likely due to background. The
resulting source count rate is $2.8 \times 10^{-3}$ counts
s$^{-1}$. Due to this low count rate and the limited spectral
resolution of the HRC-I, the spectrum of the source cannot be
constrained. We used PIMMS in order to convert the count rate to a
flux, by assuming a column density of $1.8 \times 10^{22}$ cm$^{-2}$
(as determined for Terzan 1 [Johnston et al. 1995] and assuming the
source is located in this cluster) and different spectral shapes. For
a black-body spectrum with $kT$ of 0.2--0.3 keV the unabsorbed 0.5--10
keV flux is $3 - 7 \times 10^{-13}$ \funit~(resulting in a luminosity
of $1 - 2 \times 10^{33}$ \Lunit~for a distance of 5.2 kpc) and for a
power-law spectrum with photon index of 2 the flux would be
$4.0\times10^{-13}$ \funit~($1.3\times10^{33}$ \Lunit). The possible
optical identification of this source and its nature will be further
discussed by Cody et al. (2002).

When using {\it wavdetect}, no source is detected in the {\it ROSAT}
or {\it VLA} error circles of X 1732--304.  However, when visual
inspecting the region of the {\it ROSAT} error circles
(Fig.~\ref{fig:chandra} right panel), two possible sources are
suggested by the data, one with 3 counts (at R.A = 17$^h$ 35$^m$
47.272$^s$, Dec. = --30\degr~28$'$ 55.5$''$, error $\sim$ 1$''$) and
one with only 2 counts (at R.A = 17$^h$ 35$^m$ 47.313$^s$, Dec. =
--30\degr~28$'$ 51.3$''$, error $\sim$ 1$''$).  However, the detection
of either source is statistically not significant and those sources
could be due to chance superposition of background photons. Longer
{\it Chandra} observations are needed to confirm the presence of both
sources.  If the presence of those sources can be confirmed, then
their positions are consistent (within the {\it Chandra} pointing
errors) with the {\it ROSAT} and {\it VLA} positions of X 1732--304.
However, for now, we assume that we did not detect X 1732--304 and
that less than 5 counts have been observed from it, resulting in a
count rate upper limit of $1.4\times10^{-3}$ counts s$^{-1}$. By
assuming that the quiescent spectrum of this source is very similar to
that of the other quiescent neutron star transient, we converted this
count rate limit into a flux upper limit using PIMMS. For a black-body
shaped spectrum with $kT$ of 0.2--0.3 keV and a column density of
$1.8\times10^{22}$ cm$^{-2}$, the unabsorbed 0.5--10 keV flux upper
limit would be $1.6 - 3.5 \times10^{-13}$ \funit.

Using our new {\it Chandra} result on X 1732--304, we can look back at
the {\it BeppoSAX} observation of Terzan 1 in order to investigate
whether the {\it BeppoSAX} source is X 1732--304 or an unrelated
source.  We converted (using PIMMS) the {\it BeppoSAX} LECS and MECS
count rates listed by Guainazzi et al. (1999) into predicted {\it
Chandra}/HRC-I count rates using a column density of
$1.8\times10^{22}$ cm$^{-2}$ and the power-law shaped spectrum
observed (with photon index of 2.2\pp0.6; Guainazzi et al. 1999). The
predicted HRC-I count rate is in the range 3 to 9 $\times10^{-3}$
counts s$^{-1}$, which is consistent with the count rates observed for
the 10 count source combined with the count rate upper limit on X
1732--304. Therefore, no strong evidence is available for variability
between the {\it BeppoSAX} and {\it Chandra} observations.  However, X
1732--304 is currently not the brightest X-ray source in the cluster
indicating that most if not all of the flux detected by {\it BeppoSAX}
might not have come from this source but from the 10 count source we
discovered.

The current quiescent state of X 1732--304 allows for the possible
radio identification to be verified and a search for the optical
counterpart of the source. The radio and optical observations reported
by Mart\' \i~ et al. (1998) and Ortolani et al. (1999) were performed
during times when X 1732--304 was still actively accreting. The
current lack of significant accretion in this system will most likely
also considerably have reduced its radio and optical emission.

\section{Discussion\label{section:discussion}}

We presented a short {\it Chandra}/HRC-I observation of the globular
cluster Terzan 1, known to contain the bright neutron star low-mass
X-ray binary X 1732--304. Although we detect one source with a 0.5--10
keV luminosity of $1 - 2 \times 10^{33}$ \Lunit, we could not detect X
1732--304 conclusively, with a luminosity upper limit of 0.5 -- 1
$\times10^{33}$ \Lunit~(0.5--10 keV; for a black-body shaped spectrum
with $kT$ is 0.2--0.3 keV). Brown et al. (1998) argued that the
quiescent emission of neutron star systems is due to thermal emission
from the neutron star surface and that the X-ray spectrum should be
fitted with a neutron star atmosphere model and not a black
body. Rutledge et al. (1999) showed that indeed such models can fit
the quiescent data, and that the bolometric luminosity obtained is
about twice the 0.5--10 keV luminosity. Therefore, we assume a
bolometric flux upper limit of $3 - 7 \times 10^{-13}$ \funit~($1 - 2
\times 10^{33}$ \Lunit) for X 1732--304.

Brown et al. (1998) further argued that if the quiescent luminosity
should depend on the time-averaged accretion rate of the source then a
distance independent relation can be derived between the time-averaged
flux $\langle F \rangle$ and the quiescent flux $ F_{\rm q}$ of $
F_{\rm q} \approx \langle F \rangle/135 $ (see, e.g., Rutledge et
al. 2002; but neglecting neutrino emission from the core).  The latter
can be rewritten as $\langle F \rangle = t_{\rm o} \langle F_{\rm o}
\rangle / (t_{\rm o} + t_{\rm q})$ resulting in $F_{\rm q} \approx
{t_{\rm o} \over t_{\rm o} + t_{\rm q}} \times {\langle F_{\rm o}
\rangle \over 135}$, with $\langle F_{\rm o} \rangle$ the average flux
during outburst, $t_{\rm o}$ the average time the source is in
outburst, and $t_{\rm q}$ the average time the source is in quiescence
(see also Wijnands et al. 2001).  We can estimate $\langle F_{\rm o}
\rangle$ via Figure 3 of Guainazzi et al. (1998), from which it can be
deduced that the average 2--10 keV outburst luminosity is around
10$^{36}$ \Lunit. Due to the relatively high column density towards
Terzan 1, the average bolometric luminosity can easily be a factor 5
or more higher (PIMMS indeed gives a factor $\sim$5 difference between
the 2--10 keV absorbed flux and the bolometric unabsorbed flux, using
the column density towards Terzan 1 and a power-law spectrum with
photon index of 2). Therefore, we assume a bolometric luminosity of
$5\times10^{36}$ \Lunit, resulting in a bolometric flux $\langle
F_{\rm o} \rangle$ of $1.5\times10^{-9}$ \funit. By using $F_{\rm q}$
of $<7\times10^{-13}$ \funit~ and a $t_{\it o}$ of 12 years, then the
$t_{\it q}$ is $>$180 years.

We stress that this derived lower limit is subject to large errors
because of the uncertainties in the numbers used.  For example, we
have assumed an outburst duration of 12 years for the last
outburst. However, this should be considered a lower limit because
although persistent emission from X 1732--304 was only first detected
in the mid 1980's (e.g. Skinner et al. 1987; Parmar et al. 1989), the
source was already detected through X-ray bursts in 1980 with {\it
Hakucho} (Makishima et al. 1981; Inoue et al. 1981) indicating that
the source was already actively accreting during that period. When
assuming that $t_{\it o}$ is more like 17 years then $t_{\it q}$ would
be $>250$ years.  We have also assumed that the averaged flux during
outburst and the duration of the outburst are very similar between
distinct outbursts. For X 1732--304 we cannot test those assumptions
because so far only one outburst has been observed from this source.
From other recurrent transients it is clear that arguments in favor
and against those assumptions can be made, so for simplicity, we
assume that it is true for X 1732--304. Further (i.e., longer) {\it
Chandra} observations are also needed to determine the exact quiescent
luminosity of X 1732--304 (including its quiescent spectrum) to
constrain further the time the source is inferred to be quiescent.

Assuming that all the above mentioned assumptions are valid, then X
1732--304 would be the second system after KS 1731--260 (Wijnands et
al. 2001) which has been identified as possibly having rather long
quiescent episodes. Remarkably, both systems have very long accretion
episodes and long inferred quiescent episodes (note that this might
also be true for the long-duration transient 4U 2129+47; Wijnands
2002; Nowak, Heinz, \& Begelman 2002). In contrast, the ordinary
transients which have been detected in quiescence have short outburst
episodes and relatively short quiescent ones (see, e.g., Chen,
Shrader, Livio 1997 for the behavior of ordinary transients). This
division in two groups is suggestive of the presence of a correlation
between the duration of the active episode and that of the quiescent
one. It is unclear if such a correlation can be explained in the
current disk instability models (e.g., Lasota 2001).

One possible explanation for this apparent correlation might be that
in the long-duration transients enhanced neutrino cooling occurs in
the core of their neutron stars, and in the ordinary transients only
standard cooling.  Colpi et al. (2001) suggested that when the neutron
star mass exceeds 1.65 \mzon, that this enhanced cooling occurs in the
core. If both types of transient systems have similar quiescent
episodes, the significantly higher time-averaged accretion rate in the
long-duration systems compared to the ordinary ones, will increase the
mass of the neutron stars in the long-duration transients faster than
those in the normal ones. If one assumes that the systems are roughly
of equal age, than the long-duration transients will have neutron
stars with higher masses and thus are more likely to have enhanced
neutrino emission in their neutron star cores.

Alternatively, the quiescent emission might not originate from the
neutron star surface, but might be due to some other process, such as
residual accretion or models in which the neutron star magnetic field
is highly involved (e.g., Stella et al. 1994; Menou et al. 1999;
Campana \& Stella 2000). As already discussed by Wijnands (2002), in
such models it is expected that, regardless of the outburst histories,
the quiescent emission of the different systems should be very similar
if their system parameters (i.e., orbital period; spin, mass, and
magnetic field strength of the neutron star) are very similar.  Such
assumptions are not unrealistic: from the burst oscillations during
type-I X-ray bursts, we have a good handle on the spin frequency of
the neutron star in several normal transients such as Aql X-1 (549 Hz;
Zhang et al. 1998) and 4U 1608--52 (619 Hz; Chakrabarty et al. 2000)
and in several long-duration transients KS 1731--260 (524 Hz; Smith,
Morgan, \& Bradt 1997) and MXB 1659--298 (567 Hz; Wijnands,
Strohmayer, \& Franco 2001), which are all in a very narrow
range. Moreover, the orbital periods of Aql X-1 (19 hrs; Welsh,
Robinson, \& Young 2000) and 4U 1608--52 (12.9 hrs; Wachter et al.
2002) are not extremely different from that of MXB 1659--298 (7.1 hrs;
Cominsky \& Wood 1984), which indicate similar mass transfer rates
from the companion star (see Narayan, Garcia, \& McClintock 2001 for a
discussion). Therefore, in those alternative models it is natural to
expect that the quiescent properties are very similar among the
different type of systems. The small differences in exact luminosity
can easily be explained by invoking small difference in, e.g., the
amount of residual accretion or the actual strength of the magnetic
field.

\acknowledgments

This work was supported by NASA through Chandra Postdoctoral
Fellowship grant number PF9-10010 awarded by CXC, which is operated by
SAO for NASA under contract NAS8-39073. RW thanks Tim Oosterbroek and
Matteo Guainazzi for helpful discussions about the {\it BeppoSAX} data
on Terzan 1. CH and JG thank Tom Aldcroft and Eric Schlegel for
discussions about the Chandra data.

\clearpage

\end{document}